\documentclass{article}

 \PassOptionsToPackage{numbers, compress}{natbib}

 \usepackage[preprint]{nips_2018}

\usepackage[utf8]{inputenc} 
\usepackage[T1]{fontenc}    
\usepackage{hyperref}       
\usepackage{url}            
\usepackage{booktabs}       
\usepackage{amsfonts}       
\usepackage{nicefrac}       
\usepackage{microtype}      

\usepackage[utf8]{inputenc} 
\usepackage[english]{babel} 

\usepackage{graphicx}
\usepackage{mathtools, cuted}
\usepackage{hyperref}
\usepackage{amsmath}
\usepackage{amssymb}
\usepackage{mathtools}
\usepackage{amsthm}

\newtheorem{theorem}{Theorem}[section]
\newtheorem{definition}[theorem]{Definition}

\newtheorem{lemma}[theorem]{Lemma}

\title{CM Modeling of Trajectory}

\author{
  Reza Rezaie and X. Rong Li \\
 Department of Electrical Engineering\\
 University of New Orleans\\
New Orleans, LA 70148 \\
  \texttt{rrezaie@uno.edu} and \texttt{xli@uno.edu} \\
}

\begin{document}

\maketitle

\begin{abstract}
Information about the waypoints of a moving object, e.g., an airliner in an air traffic control (ATC) problem, should be considered in trajectory modeling and prediction. Due to the ATC regulations, trajectory design criteria, and restricted motion capability of airliners there are long range dependencies in trajectories of airliners. Waypoint information can be used for modeling such dependencies in trajectories. This paper proposes a conditionally Markov (CM) sequence for modeling trajectories passing by waypoints. A dynamic model governing the proposed sequence is obtained. Filtering and trajectory prediction formulations are presented. The use of the proposed sequence for modeling trajectories with waypoints is justified.

\end{abstract}


\section{Introduction}

Markov processes have been widely used for modeling random phenomena. A Markov process has two main components: an initial distribution and an evolution law. However, for some problems Markov processes are not adequate. Then, sometimes a higher-order (e.g., second-order) Markov process is used. But such a model does not fit some phenomena well, for example, a time-varying phenomenon with some information available about its future. An example is the problem of predicting the trajectory of an airliner in the presence of waypoint information. The Markov process does not fit such a problem because the future distribution of a Markov process is completely determined by its initial distribution and evolution law. 

Trajectory modeling and prediction in the presence of an intent or a destination has been studied in the literature. \cite{Hwang0}--\cite{Krozel2} presented some intent-based trajectory prediction approaches for air traffic control (ATC). Some trajectory prediction approaches were presented in \cite{Hwang0}--\cite{Hwang2} based on hybrid estimation aided by intent information. In \cite{Louis}, the interacting multiple model (IMM) approach was used for trajectory prediction, where a higher weight was assigned to the model with the closest heading towards the waypoint. \cite{Krozel} presented an approach for trajectory prediction using an inferred intent based on a database. In \cite{Krozel2}, the use of waypoint information for trajectory prediction in ATC was discussed. Ship trajectories were modeled by a Gauss-Markov model in \cite{Ship}, where predictive information was incorporated. After quantizing the state space, \cite{Fanas1}--\cite{White_Tracking1} used finite-state reciprocal sequences for intent inference and a generalization of reciprocal sequences for trajectory modeling. A problem with quantized state space is the complexity of the corresponding estimation algorithms. So, the complexity of the algorithms used in \cite{Fanas1}--\cite{White_Tracking1} was also addressed. The Gaussian counterpart of the generalized reciprocal sequence defined in \cite{White_Tracking1} was studied in \cite{White_Tracking2}. \cite{S_1}--\cite{S_2} used bridging distributions for the purpose of intent inference, for example, in selecting an icon on an in-vehicle interactive display. A CM sequence was used in \cite{DD_Conf} for trajectory modeling with destination information. A systematic framework for modeling trajectories with waypoints is desired.

Inspired by \cite{Mehr}, a class of CM sequences, called $CM_L$, was defined, modeled, and characterized in \cite{CM_Part_I_Conf}. A second-order nearest-neighbor dynamic model driven by locally correlated dynamic noise was presented in \cite{Levy_Dynamic} for the nonsingular Gaussian (NG) reciprocal sequence. As special CM sequences, in \cite{CM_Part_II_A_Conf}--\cite{CM_Explicitly} NG reciprocal sequences were studied from the CM viewpoint. Also, some dynamic models with white dynamic noise were presented for the NG reciprocal sequence. 

Consider stochastic sequences defined over $[0,N]=\lbrace 0,1,\ldots,N \rbrace$. For convenience, let the index be time. A sequence is Markov if and only if (iff) conditioned on the state at any time $k$, the subsequences before and after $k$ are independent. A sequence is reciprocal iff conditioned on the states at any two times $k_1$ and $k_2$, the subsequences inside and outside the interval $[k_1,k_2]$ are independent. In other words, ``inside" and ``outside" are independent given the boundaries. A sequence is $CM_L$ iff conditioned on the state at time $N$, the sequence is Markov over $[0,N-1]$. 

The main components for modeling trajectories without any future information (no information about future waypoints or destination) are an origin and an evolution law. Because a Markov process is determined by its initial distribution and evolution law, Markov processes can model such trajectories. The main components for modeling trajectories with destination information (called \textit{destination-directed trajectories}) are an origin, an evolution law, and a destination. The main elements of the $CM_L$ sequence are a joint endpoint distribution (i.e., an initial distribution and conditioned on it a final distribution) and a Markov-like evolution law. The $CM_L$ sequence can model the main components of destination-directed trajectories \cite{DD_Conf}, but not trajectories passing by waypoints. Due to the ATC regulations, trajectory design criteria, restricted motion capability of airliners, and the ATC trajectory repeatability and predictability requirements \cite{ATC} there are long range dependencies in trajectories of airliners. Waypoint information can be used for modeling such dependencies in trajectories. Assume an airliner broadcasts its next waypoint by the time it passes its current waypoint. Trajectories start from an origin, pass the waypoints, and end at a destination (which can be seen as the last waypoint). This paper proposes a CM sequence for modeling such trajectories. Properties of the proposed CM sequence for modeling trajectories with waypoints are discussed. The corresponding dynamic model, filter, and trajectory predictor are obtained.
  
The paper is organized as follows. In Section \ref{DD_Modeling}, modeling of destination-directed trajectories by the $CM_L$ sequence is discussed. Then, a CM sequence is presented for modeling trajectories passing by waypoints. Also, the corresponding dynamic model is obtained. In Section \ref{Filtering_Prediction}, the filter and the trajectory predictor are presented. In Section \ref{Simulation}, the presented model is simulated for trajectory prediction with waypoints. Section \ref{Summary}, includes conclusions.

\section{Trajectory Modeling Using CM Sequences}\label{DD_Modeling}

The following notation is used for time intervals and stochastic sequences:
\begin{align*}
[i,j]& \triangleq \lbrace i,i+1,\ldots ,j-1,j \rbrace\\
[x_k]_{i}^{j} & \triangleq \lbrace x_k, k \in [i,j] \rbrace\\
[x_k] & \triangleq [x_k]_{0}^{N}
\end{align*}
where $k$ in $[x_k]_i^j$ is a dummy variable. Also, ZMNG and NG stands for ``zero-mean nonsingular Gaussian" and ``nonsingular Gaussian", respectively. We consider sequences defined over $[0,N]$. $F(\cdot | \cdot)$ denotes a conditional cumulative distribution function (CDF).

\subsection{$CM_L$ Sequences for Destination-Directed Trajectory Modeling}

We review the definition and a dynamic model of the $CM_L$ sequence for destination-directed trajectory modeling \cite{DD_Conf}, \cite{CM_Part_I_Conf}, \cite{CM_Part_II_A_Conf}--\cite{CM_Part_II_B_Conf}. 

\begin{definition}\label{Markov} 
$[x_k]$ is Markov if $\forall j,k \in [0,N], j<k$,
\begin{align}
F(x_k|[x_i]_0^j) =F(x_k|x_{j})\label{Markov_Def}
\end{align}

\end{definition}

Sample paths of some Markov sequences can be used for modeling trajectories without waypoint or destination information. For example, a nearly constant velocity, acceleration, or turn motion model (with white noise) is a Markov model. 
\begin{lemma}\label{Model_Dynamic_Proposition}
A ZMNG $[x_k]$ with covariance function $C_{l_1,l_2}$ is Markov iff its evolution is governed by
\begin{align}
x_k=M_{k,k-1}x_{k-1}+e^M_{k}, k \in [1,N], \quad x_0=e^M_0 \label{Markov_Model}
\end{align}
where $[e^M_k]$ is a zero-mean white NG sequence with covariances $M_k$.
\end{lemma}
     
The Markov sequence is not powerful enough for modeling an origin, an evolution law, and a destination. Since the future distribution of a Markov sequence is determined by its initial distribution and evolution law, it is not powerful enough to model future information. A more general class of stochastic sequences ($CM_L$ sequences) was used in \cite{DD_Conf} for modeling trajectories with destination information (destination-directed trajectories). It can be justified as follows. Let destination-directed trajectories be modeled as the sample paths of a sequence $[x_k]$. Since the destination of the trajectories (i.e., density of $x_N$) is known, the evolution law can be modeled as a conditional density given the destination $x_N$. This conditional density is chosen to be a Markov density, i.e., $[x_k]_0^{N-1}$ being Markov conditioned on $x_N$. This evolution law, which is simple and desirable for modeling destination-directed trajectories, corresponds to the $CM_L$ sequence defined as follows \cite{CM_Part_I_Conf}.  
\begin{definition}\label{CML_Like}
$[x_k]$ is $CM_L$ if $\forall j, k \in [0,N-1], j<k$,
\begin{align}
F(x_k|[x_i]_0^j,x_N) =F(x_k|x_j,x_N)\label{CML_Def}
\end{align}

\end{definition}

In other words, $[x_k]$ is $CM_L$ iff conditioned on $x_{j}$ and $x_{N}$ ($\forall j \in [1,N-2]$), the subsequences $[x_k]_{j+1}^{N-1}$ and $[x_k]_{0}^{j-1}$ are independent. 

A dynamic model for the evolution of the $CM_L$ sequence, called a $CM_L$ model, is as follows \cite{CM_Part_I_Conf}.
\begin{theorem}\label{CML_Forward_Dynamic_Proposition}
A ZMNG $[x_k]$ with convariance function $C_{l_1,l_2}$ is $CM_L$ iff its evolution is governed by
\begin{align}\label{CML_Model}
x_k=G_{k,k-1}x_{k-1} + G_{k,N}x_N+e_k, \quad k \in [1,N-1]
\end{align}
where $[e_k]$ is a zero-mean white NG sequence with covariances $G_k$, and either boundary conditions
\begin{align}
x_0=e_0, \quad x_N=G_{N,0}x_0+e_N\label{CML_BC1}\\
x_N=e_N, \quad x_0=G_{0,N}x_N+e_0\label{CML_BC2}
\end{align}

\end{theorem}

A non-zero-mean Gaussian sequence is $CM_L$ (or Markov) iff its zero-mean part follows the dynamic model of Theorem \ref{CML_Forward_Dynamic_Proposition} (or Lemma \ref{Model_Dynamic_Proposition}). The same is true for the sequence defined later. Therefore, for simplicity and brevity, we consider zero-mean sequences, but in simulations non-zero-mean sequences are used.

An approach for the $CM_L$ model parameter design for modeling destination-directed trajectories is as follows \cite{DD_Conf}. Such trajectories can be modeled by combining (superimposition of) two key assumptions: (i) the moving object follows a Markov model $\eqref{Markov_Model}$ (e.g., a nearly constant velocity model) without considering the destination information, and (ii) the destination density is known (which can differ from the destination density of the Markov model in (i)). Let $[s_k]$ be a Markov sequence governed by $\eqref{Markov_Model}$ (e.g., a nearly constant velocity model). Since every Markov sequence is $CM_L$, $[s_k]$ can also obey a $CM_L$ model as
 \begin{align}
s_k&=G_{k,k-1}s_{k-1}+G_{k,N}s_N+e^s_k, k \in [1,N-1]\label{CML_Dynamic_for_Markov}\\
s_N&=e^{s}_N, \quad s_0=G^{s}_{0,N}s_N+e^{s}_0\label{CML_R_FQ_BC2}
\end{align}
where $[e^s_k]$ is a zero-mean white NG sequence with covariances $G_k, k \in [1,N-1]$, $G^s_0$, and $G^s_N$.

Parameters of $\eqref{CML_Dynamic_for_Markov}$ can be obtained as follows. By $\eqref{Markov_Model}$, we have $p(s_k|s_{k-1})=\mathcal{N}(s_k;M_{k,k-1}s_{k-1},$ $M_{k})$. Since $[s_k]$ is Markov, we have ($ k \in [1,N-1]$)
\begin{align}
p(s_k|s_{k-1},s_N)&=\frac{p(s_k|s_{k-1})p(s_N|s_k)}{p(s_N|s_{k-1})}\label{CML_Reciprocal_Transition}\\
&=\mathcal{N}(s_k;G_{k,k-1}s_{k-1}+G_{k,N}s_N;G_k)\nonumber
\end{align}
where $G_{k,k-1}$, $G_{k,N}$, and $G_k$ are obtained as
\begin{align}
G_{k,k-1}&=M_{k,k-1}-G_{k,N}M_{N|k-1} \label{CML_Choice_1}\\
G_{k,N}&=G_kM_{N|k}'C_{N|k}^{-1} \label{CML_Choice_2}\\
G_k&=(M_{k}^{-1}+M_{N|k}'C_{N|k}^{-1}M_{N|k})^{-1}\label{CML_Choice_3}\\
M_{N|k}&=M_{N,N-1}\cdots M_{k+1,k}, k \in [1,N-1], M_{N|N}=I\nonumber\\
C_{N|k}&=\sum _{n=k}^{N-1}M_{N|n+1}M_{n+1}M_{N|n+1}', k \in [1,N-1]\nonumber\\
p(s_N|s_i)&=\mathcal{N}(s_N;M_{N|i}s_{i},C_{N|i}), i \in [0,N-1]\nonumber
\end{align}
and $M_{k,k-1}, M_k, k \in [1,N]$, are parameters of $\eqref{Markov_Model}$.

Now, we construct a sequence $[x_k]$ governed by
\begin{align}
x_k&=G_{k,k-1}x_{k-1}+G_{k,N}x_N+e_k, k \in [1,N-1]\label{CML_Dynamic_for_Markov_x}\\
x_N&=e_N, \quad x_0=G_{0,N}x_N+e_0\label{CML_R_FQ_BC2_x}
\end{align}
where $[e_k]$ is a zero-mean white NG sequence with covariances $G_k, k \in [1,N-1], G_0$, and $G_N$. Note that $\eqref{CML_Dynamic_for_Markov_x}$ and $\eqref{CML_Dynamic_for_Markov}$ have the same parameters ($G_{k,k-1}, G_{k,N},G_k, k \in [1,N-1]$), but parameters of $\eqref{CML_R_FQ_BC2_x}$ ($G_{0,N},G_0,G_N$) and parameters of $\eqref{CML_R_FQ_BC2}$ ($G^s_{0,N},G^s_0,G^s_N$) are different. Parameters of $\eqref{CML_R_FQ_BC2_x}$ ($G_{0,N},G_0,G_N$) can be chosen arbitrarily (i.e. $G_{0,N}$ can be any matrix with suitable dimension, and $G_0$ and $G_N$ any positive definite matrix with suitable dimension). Thus, $[x_k]$ can have any joint density of $x_0$ and $x_N$. So, $[s_k]$ and $[x_k]$ have the same $CM_L$ model ($\eqref{CML_Dynamic_for_Markov}$ and $\eqref{CML_Dynamic_for_Markov_x}$) (in other words, the same transition density $\eqref{CML_Reciprocal_Transition}$), but $[x_k]$ can have any joint endpoint density. It means any origin and destination of $[x_k]$ can be so modeled. Therefore, combining assumptions (i) and (ii) above naturally leads to the $CM_L$ sequence $[x_k]$ whose $CM_L$ model is the same as that of $[s_k]$ while the former can model any origin and destination. 

Reciprocal sequences are special $CM_L$ sequences. 
\begin{definition}\label{Reciprocal} 
$[x_k]$ is reciprocal if $\forall k_1, k, k_2 \in [0,N]$, $k_1<k<k_2$,
\begin{align}
F(x_k|[x_i]_0^{k_1},[x_i]_{k_2}^{N})=F(x_k|x_{k_1},x_{k_2})\label{Reciprocal_Def}
\end{align}
\end{definition}

The $CM_L$ model $\eqref{CML_Dynamic_for_Markov_x}$ with $\eqref{CML_Choice_1}$--$\eqref{CML_Choice_3}$ is called a $CM_L$ model induced by a Markov model. By Theorem \ref{CML_R_Dynamic_FQ_Proposition} below, such a $CM_L$ model governs a reciprocal sequence (so it is called a reciprocal $CM_L$ model \cite{CM_Part_II_A_Conf}). Also, Theorem \ref{CML_R_Dynamic_FQ_Proposition} shows that every reciprocal $CM_L$ model can be \textit{induced} by a Markov model following the above approach \cite{CM_Part_II_B_Conf}.
\begin{theorem}\label{CML_R_Dynamic_FQ_Proposition} 
A ZMNG $[x_k]$ is reciprocal iff it obeys $\eqref{CML_Model}$ and $\eqref{CML_BC2}$, where $(G_{k,k-1},G_{k,N},G_k)$, $k \in [1,N-1]$, are given by $\eqref{CML_Choice_1}$--$\eqref{CML_Choice_3}$, $M_{k,k-1}$, $ k \in [1,N]$, are square matrices, and $M_k$, $k \in [1,N]$, are positive definite matrices with the dimension of $x_k$.

\end{theorem}

A non-zero-mean Gaussian $CM_L$ sequence for modeling destination-directed trajectories is as follows. Let $\mu _0$ ($\mu _N$) and $C_0$ ($C_N$) be the mean and covariance of the origin (destination) distribution. Also, let $C_{0,N}$ be the cross-covariance of the states at the origin and the destination. So, $x_N \sim \mathcal{N}(\mu _N, C_N)$. Then, $x_0=\mu _0+G_{0,N}(x_N-\mu _N)+e_0$, where $G_{0,N}=C_{0,N}C_N^{-1}$ and $\text{Cov}(e_0)=C_0-C_{0,N}C_N^{-1}$ $\cdot (C_{0,N})'$. In addition, the state evolution for $k \in [1,N-1]$ is governed by $\eqref{CML_Model}$.

\subsection{A CM Sequence for Trajectory Modeling with Waypoint Information}

Consider trajectories of airliners in ATC. An airliner passes several waypoints before reaching the destination. The waypoint information (about the location of the waypoint and the time at which the airliner should pass the waypoint) is broadcast ahead of time by the airliner. Let $N_n$ denote the time for the $n$th waypoint (the time at which the airliner should pass the waypoint). Assume the destination is not known. By time $N_n$, the airliner broadcasts its next waypoint (the $(n+1)$th waypoint) information. The main elements for trajectory modeling in this problem are consecutive waypoints and the motion between them. A simple model capable of describing these main elements is desirable. 

The states $x_{j}$ ($j \in [N_n,N_{n+1}-1]$) and $x_{N_{n+1}}$ together can provide reasonable information about the past (the time before $j$) and the intent of an airliner in order to model the trajectory between $j$ and $N_{n+1}$. Therefore, given $x_j$ and $x_{N_{n+1}}$, it is assumed that the trajectories over $[j,N_{n+1}]$ and before $j$ are independent. On the other hand, the waypoint sequence is a rough (grand scale) description of the trajectory. It is assumed that given the state at the $n$th waypoint (i.e., $x_{N_n}$), the states at later waypoints (i.e., $x_{N_{q}}, q>n$) are independent of the states at earlier waypoints (i.e., $x_{N_{q}}, q<n$). In addition, airliners usually follow their flight plan, i.e., satisfy the waypoint requirements. It means the state at a waypoint ($x_{N_n}$) can represent the state of the trajectory before (and especially close to) the waypoint. Thus, it is assumed that given $x_{N_n}$, the states $x_{N_{q}}, q>n$ are independent of all states before $x_{N_n}$. The above assumptions seem reasonable for trajectories passing waypoints.  

In the following, a CM sequence for trajectory modeling in the above problem is defined.  
\begin{definition}\label{CM_Def}
$[x_k]$ is a stochastic sequence, where
\begin{itemize}
\item[(i)] $\forall n \in$ $ [1,m]$ and $\forall j,k$, $0=N_0<N_1<\cdots < N_{n-1} \leq j <k< N_n<N_{n+1} $ $ < \cdots $ $ < N_{m}=N$
\begin{align}
F(x_k |[x_i]_0^{j},x_{N_n})=F(x_k |x_j,x_{N_n})
\label{CM_Def_1}
\end{align}

\item[(ii)] $\forall n \in [1,m]$ and $\forall h<n$
\begin{align}
&F(x_{N_n}|[x_i]_0^{N_h})=F(x_{N_n}|x_{N_h}) \label{CM_Def_2}
\end{align}

\end{itemize}
\end{definition}

By Definition \ref{CM_Def}, $[x_k]$ is $CM_L$ over $[N_{n-1},N_n]$. Conditioned on $x_{N_n}$ and $x_j$, $x_k, k \in [j+1,N_n-1]$ is independent of $[x_k]_0^{j-1}$. Also, given $x_{N_h}$, $x_{N_n}$ is independent of $[x_k]_0^{N_h-1}$. 

A stochastic sequence $[x_k]$ can be generated in many different ways. Let the density function of $[x_k]$ exist and be denoted by $p([x_k])$. Then, sample paths of $[x_k]$ can be generated in time order (i.e., $x_0, x_1, \ldots, x_N$) according to the following decomposition
\begin{align}
&p([x_k])=\nonumber\\
&p(x_N|[x_k]_0^{N-1})p(x_{N-1}|[x_k]_0^{N-2})\cdots p(x_1|x_0)p(x_0)\label{x_generation}
\end{align}
Based on $\eqref{x_generation}$, first $x_0$ is generated from $p(x_0)$. Then, given $x_0$, $x_1$ is generated from $p(x_1|x_0)$, and so on. 

Based on its properties, a simple way for sample-path generation of the sequence $[x_k]$ with Definition \ref{CM_Def} is as follows. First, $x_0$ is generated from $p(x_0)$ and $x_{N_1}$ is generated from $p(x_{N_1}|x_0)$. Then, given $x_0$ and $x_{N_1}$, $[x_k]_1^{N_1-1}$ are generated in time order, i.e., $x_1 \sim p(x_1|x_0,x_{N_1})$, then $x_2 \sim p(x_2|x_1,x_{N_1})$, and so on. Then, given $[x_i]_0^{N_1}$, $x_{N_2}$ is generated from $p(x_{N_2}|x_{N_1})$. Generation of $[x_k]_{N_1+1}^{N_2-1}$ is in time order similar to that of $[x_k]_{1}^{N_1-1}$. This approach is used until the end of the sequence. The dynamic model of Theorem \ref{CM_Dynamic} below clarifies this approach. 

Before presenting the dynamic model, we have a lemma.

\begin{lemma}\label{E_CM}
A Gaussian sequence $[x_k]$ follows Definition \ref{CM_Def} iff
\begin{itemize}

\item[(i)] $\forall n \in$ $ [1,m]$ and $\forall j, k$, $0=N_0<N_1<\cdots < N_{n-1} \leq j<k < N_n<N_{n+1} $ $ < \cdots $ $ < N_{m}=N$
\begin{align}
E[x_k |[x_i]_0^{j},x_{N_n}]=E[x_k |x_j,x_{N_n}]
\label{E_CM_1}
\end{align}

\item[(ii)] $\forall n \in [1,m]$ and $\forall h<n$
\begin{align}
&E[x_{N_n}|[x_i]_0^{N_h}]=E[x_{N_n}|x_{N_h}] \label{E_CM_2}
\end{align}

\end{itemize}
\end{lemma}

A dynamic model governing a Gaussian sequence with Definition \ref{CM_Def} is obtained using Lemma \ref{E_CM}.
\begin{theorem}\label{CM_Dynamic}
A ZMNG sequence $[x_k]$ with covariance function $C_{l_1,l_2}, l_1,l_2 \in [0,N]$ follows Definition \ref{CM_Def} iff $ \forall k \in [N_{n-1}+1,N_n-1]$ and $\forall n \in [1,m]$,
\begin{align}
x_{k} &=G_{k,k-1}x_{k-1}+G_{k,N_n}x_{N_n}+e_{k} \label{D1}\\
x_{N_n} &= G_{N_n,N_{n-1}}x_{N_{n-1}}+e_{N_n}\label{D2}
\end{align}
where $[e_k]_1^N$ is a zero-mean white Gaussian sequence with nonsingular covariances $G_k$, uncorrelated with $x_0$ with nonsingular covariance $G_0$. 

\end{theorem}

A non-zero-mean Gaussian sequence satisfies Definition \ref{CM_Def} iff its zero-mean part is governed by the model in Theorem \ref{CM_Dynamic}.

The use of Definition \ref{CM_Def} for trajectory modeling with waypoints is discussed. Let the trajectories be modeled by $[x_k]$ following Definition \ref{CM_Def}. Similar to destination-directed trajectories, the subsequence $[x_k]_{N_{n-1}}^{N_n}$ is governed by a $CM_L$ model induced by a Markov model (Theorem \ref{CML_R_Dynamic_FQ_Proposition}). So, parameters of $\eqref{D1}$ are given by $\eqref{CML_Choice_1}$--$\eqref{CML_Choice_3}$ (see Section \ref{Simulation}). By time $N_{n-1}$ the next waypoint information is available. It means the position mean of $x_{N_n}$ (and potentially other information, e.g., turn rate, navigation accuracy) is given. The remaining information corresponding to the next waypoint (i.e., the velocity mean at the waypoint, the covariance of the state at the waypoint $C_{N_{n}}$, and the cross-covariance of the states at two consecutive waypoints $C_{N_{n},N_{n-1}}$) can be learned in advance based on a set of trajectories or can be designed. The impact of any mismatch in these parameters ($\mu _{N_{n}}$, $C_{N_{n}}$, and $C_{N_{n},N_{n-1}}$) is studied in section \ref{Simulation}. Parameters of $\eqref{D2}$ are obtained using the covariance of the jointly Gaussian density of $x_{N_{n-1}}$ and $x_{N_{n}}$ as follows ($n >0$):
\begin{align*}
G_{N_n,N_{n-1}}&=C_{N_n,N_{n-1}}(C_{N_n})^{-1}\\
G_{N_n}&=C_{N_n}-C_{N_n,N_{n-1}}(C_{N_n})^{-1}(C_{N_n,N_{n-1}})'
\end{align*}
where in $\eqref{D2}$, $G_0=C_0$. In section \ref{Simulation}, a non-zero-mean Gaussian CM sequence with Definition \ref{CM_Def} is used for trajectory modeling and prediction. The corresponding CM sequence is governed by the dynamic model in Theorem \ref{CM_Dynamic}, where instead of $\eqref{D2}$ we have 
\begin{align}
x_{N_n}=\mu _{N_n} + G_{N_n,N_{n-1}}(x_{N_{n-1}} - \mu _{N_{n-1}}) + e_{N_n}\label{D3}
\end{align}

\section{Filtering and Prediction}\label{Filtering_Prediction}

\subsection{Filtering}\label{Filtering}

Consider model in Theorem \ref{CM_Dynamic}, where instead of $\eqref{D2}$ we have $\eqref{D3}$, and measurement model
\begin{align}\label{measurement}
z_k=H_kx_k+v_k, \quad k \in [1,N]
\end{align}
where $[v_k]_1^N$ is zero-mean white Gaussian noise with $\text{Cov}(v_k)=R_k$, uncorrelated with $x_0$ and $[e_k]_1^{N}$. We want to obtain $\hat{x}_k=E[x_k|z^k]$ and its mean square error (MSE) matrix given all measurements from the start to time $k$, denoted by $z^{k}=\lbrace z_1, z_2, \ldots, z_k\rbrace$.

For $k \in [0,N_1-1]$, let $y_k=[x_k' , x_{N_1}']$. Given the jointly Gaussian density $\mathcal{N}(y_0;\mu^y _0, C^y_0)$, the minimum MSE (MMSE) estimate of $y_0$ and its MSE matrix are $\hat{y}_0=\mu ^y_0$ and $\Sigma _0=C^y_0$. For $n=1$, $\eqref{D1}$ can be written as 
\begin{align}\label{CML_2}
y_k=G^y_{k,k-1}y_{k-1}+e^y_{k-1}, \quad  k \in [1,N_1-1]
\end{align}
where
\begin{align}
G^y_{k,k-1}&=\left[\begin{array}{cc}
G_{k,k-1} & G_{k,N_1}\\
0 & I
\end{array}\right]\\
e^y_k&=\left[\begin{array}{c}
e_{k+1}\\
0
\end{array}\right], \quad 
G^y_k=\text{Cov}(e^y_k)=\left[
\begin{array}{cc}
G_{k+1} & 0\\
0 & 0
\end{array}\right]
\end{align}
In addition, $\eqref{measurement}$ is written as
\begin{align}\label{measurement_2}
z_k=H^y_ky_k+v_k, \quad k \in [1,N]
\end{align}
where $H^y_k=[H_k ,0]$. Based on $\eqref{CML_2}$ and $\eqref{measurement_2}$, the MMSE estimator and its MSE matrix are
\begin{align}
\hat{y}_k&=E[y_k|z^k]\nonumber \\
&=\hat{y}_{k|k-1}+C_{y_k,z_k}C_{z_k}^{-1}\Big(z_k-H_k^y\hat{y}_{k|k-1}\Big)\label{y_h}\\
\Sigma _k&=E[(y_k-\hat{y}_k)(y_k-\hat{y}_k)']\nonumber\\
&=\Sigma_{k|k-1}-C_{y_k,z_k}C_{z_k}^{-1}(C_{y_k,z_k})'\label{P_y_h}
\end{align}
where 
\begin{align*}
\hat{y}_{k|k-1}&= G^y_{k,k-1}\hat{y}_{k-1}\\
\Sigma_{k|k-1}&= G^y_{k,k-1}\Sigma _{k-1}(G^y_{k,k-1})'+G^y_{k-1}\\
C_{y_k,z_k}&=\Sigma _{k|k-1}(H_k^y)'\\
C_{z_k}&=H_k^y\Sigma _{k|k-1}(H_k^y)'+R_k
\end{align*}
and the estimate of $x_k$ and its MSE are ($k \in [1,N_1-1]$)
\begin{align*}
\hat{x}_k&=[I ,0]\hat{y}_k\\
P_k&=[I ,0]\Sigma _k [I , 0]'
\end{align*}
Given $\hat{y}_{N_1-1}$ and $\Sigma _{N_1-1}$, 
\begin{align*}
\hat{x}_{N_1|N_1-1}&=[0 , I]\hat{y}_{N_1-1}\\
P_{N_1|N_1-1}&=[0 , I]\Sigma _{N_1-1} [0 , I]'
\end{align*}
where $\hat{x}_{N_1|N_1-1}$ is the estimate of $x_{N_1}$ given all measurements up to time $N_1-1$, and $P_{N_1|N_1-1}$ is its MSE matrix. Then, given $z_{N_1}$, $\hat{x}_{N_1|N_1-1}$ and $P_{N_1|N_1-1}$ are updated as
\begin{align}
\hat{x}_{N_1}&=\hat{x}_{N_1|N_1-1}+C_{x_{N_1},z_{N_1}}C_{z_{N_1}}^{-1}(z_{N_1}-H_{N_1}\hat{x}_{N_1|N_1-1})\label{N_filter_m}\\
P_{N_1}&=P_{N_1|N_1-1}-C_{x_{N_1},z_{N_1}}C_{z_{N_1}}^{-1}(C_{x_{N_1},z_{N_1}})'\label{N_filter_C}
\end{align}
where 
\begin{align*}
C_{x_{N_1},z_{N_1}}&=P_{N_1|N_1-1}(H_{N_1})'\\
C_{z_{N_1}}&=H_{N-1}P_{N_1|N_1-1}(H_{N_1})'+R_{N_1}
\end{align*}

$p(x_{N_1},x_{N_2}|z^{N_1})$ is the posterior jointly Gaussian density of $x_{N_1}$ and $x_{N_2}$. To estimate $x_{N_1+1}$ (based on $\eqref{D1}$), we need to calculate $p(x_{N_1},x_{N_2}|z^{N_1})$ with the following conditional mean and conditional covariance
\begin{align}\left[
\begin{array}{c}
E[x_{N_{1}}|z^{N_{1}}]\\
E[x_{N_2}|z^{N_{1}}]
\end{array}\right],
\left[
\begin{array}{cc}
C_{N_{1}|N_{1}} & C_{N_{1},N_2|N_{1}}\\
C_{N_{1},N_2|N_{1}}' & C_{N_2|N_{1}}
\end{array}\right]\label{Cnd}
\end{align}
where we have $C_{k_1|k}=\text{Cov}(x_{k_1}|z^{k})$ and $C_{k_1,k_2|k}=\text{Cov}(x_{k_1},x_{k_2}|z^{k})$. We already have $E[x_{N_{1}}|z^{N_{1}}]=\hat{x}_{N_1}$ and $C_{N_{1}|N_{1}}=P_{N_1}$. $E[x_{N_2}|z^{N_{1}}]$ and $C_{N_2|N_1}$ are calculated as follows. By $\eqref{D3}$ and the whiteness and the uncorrelatedness of $[v_k]_1^N$, $[e_k]_1^N$, and $x_0$, we have 
\begin{align}
p(x_{N_2}|x_{N_{1}})=p(x_{N_2}|x_{N_{1}},z^{N_1})\label{Tm}
\end{align}
Thus,
\begin{align}
p(x_{N_2}|z^{N_{1}})&=\int p(x_{N_2}|x_{N_{1}})p(x_{N_{1}}|z^{N_{1}})dx_{N_{1}}\label{Tm1}
\end{align}
Given $\mu _{N_2}$, $C_{N_2}$, and $C_{N_2,N_1}$, based on $\eqref{Tm1}$, we have
\begin{align*}
E[x_{N_2}|z^{N_1}]&=\mu _{N_2} + G_{N_2,N_1}(\hat{x}_{N_1}-\mu _{N_1})\\
C_{N_2|N_1}&=G_{N_2}+G_{N_2,N_1}P_{N_1}(G_{N_2,N_1})'
\end{align*}
where $G_{N_2,N_1}=C_{N_2,N_1}(C_{N_1})^{-1}$ and $G_{N_2}=C_{N_2}-C_{N_2,N_1}$ $\cdot(C_{N_1})^{-1}(C_{N_2,N_1})'$. 

$C_{N_1,N_2|N_1}$ is calculated as follows. Consider the mean and covariance of $p(x_{N_1},x_{N_2}|z^{N_1})$ given by $\eqref{Cnd}$. We have
\begin{align}
E[x_{N_2}|x_{N_1}&,z^{N_1}]=E[x_{N_2}|z^{N_1}]\nonumber\\
&+C_{N_2,N_1|N_1}(C_{N_1|N_1})^{-1}(x_{N_1}-E[x_{N_1}|z^{N_1}])\nonumber\\
&=\mu _{N_2} + C_{N_2,N_1}(C_{N_1})^{-1}(\hat{x}_{N_1}-\mu _{N_1})\nonumber\\
&+C_{N_2,N_1|N_1}(C_{N_1|N_1})^{-1}(x_{N_1}-\hat{x}_{N_1})\label{C1}
\end{align}
Also,
\begin{align}
E[x_{N_2}|x_{N_1}]&=\mu _{N_2}+C_{N_2,N_1}(C_{N_1})^{-1}(x_{N_1}-\mu _{N_1})\label{C2}
\end{align}
By $\eqref{Tm}$, we have
\begin{align}
E[x_{N_2}|x_{N_1}]=E[x_{N_2}|x_{N_1},z^{N_1}]\label{Cm}
\end{align}
Substituting $\eqref{C1}$--$\eqref{C2}$ into $\eqref{Cm}$, after some manipulation, yields
\begin{align}
(C_{N_2,N_1}(C_{N_1})^{-1}-C_{N_2,N_1|N_1}(C_{N_1|N_1})^{-1})(x_{N_1}-\hat{x}_{N_1})=0\label{Eq}
\end{align}
$\eqref{Eq}$ holds for every\footnote{Actually, $\eqref{Cm}$ holds almost surely, but we consider a regular version of the conditional expectations \cite{Loeve}, where we have $\eqref{Eq}$ for every $x_{N_1}-\hat{x}_{N_1} \in \Re^d$.} $x_{N_1}-\hat{x}_{N_1} \in \Re^d$ (where $d$ is the dimension of the state vector $x_k$), i.e., $\Re^d$ is the null space and thus
\begin{align*}
G_{N_2,N_1}=C_{N_2,N_1|N_1}(C_{N_1|N_1})^{-1}
\end{align*}
which results in
\begin{align}
C_{N_2,N_1|N_1}=G_{N_2,N_1}C_{N_1|N_1}=(C_{N_1,N_2|N_1})'\label{C_N2_N1|z}
\end{align}
where $C_{N_1|N_1}=P_{N_1}$.

Given $p(x_{N_1},x_{N_2}|z^{N_1})$, filtering for $k \in [N_1+1,N_2-1]$ (which is similar to the filtering for $k \in [1,N_1-1]$) is based on the following model for $y_k=[x_k' , x_{N_2}']'$:
\begin{align*}
y_k=
\left[
\begin{array}{cc}
G_{k,k-1} & G_{k,N_2}\\
0 & I
\end{array}\right] y_{k-1}+
e^y_{k-1}
\end{align*}

Similarly, given $p(x_{N_{n-1}}|z^{N_{n-1}}), n>1$, we have
\begin{align*}
E[x_{N_n}|z^{N_{n-1}}]&=\mu _{N_n}+G_{N_n,N_{n-1}}(\hat{x}_{N_{n-1}}-\mu _{N_{n-1}})\\
C_{N_n|N_{n-1}}&=G_{N_n}+G_{N_n,N_{n-1}}P_{N_{n-1}}(G_{N_n,N_{n-1}})'\\
C_{N_n,N_{n-1}|N_{n-1}}&=G_{N_n,N_{n-1}}C_{N_{n-1}|N_{n-1}}\\
&=(C_{N_{n-1},N_n|N_{n-1}})'
\end{align*}
where $G_{N_n,N_{n-1}}=C_{N_n,N_{n-1}}(C_{N_{n-1}})^{-1}$ and $G_{N_n}=C_{N_n}$ $-C_{N_n,N_{n-1}}(C_{N_{n-1}})^{-1}(C_{N_n,N_{n-1}})'$. 

Therefore, $p(x_{N_{n-1}},x_{N_n}|z^{N_{n-1}})$ is available. Then, for $k \in [N_{n-1}+1,N_n-1]$, filtering is performed using the following model for $y_k=[x_k',x_{N_n}']'$:
\begin{align}
y_k=
\left[
\begin{array}{cc}
G_{k,k-1} & G_{k,N_n}\\
0 & I
\end{array}\right] y_{k-1}+
e^y_{k-1}\label{CM_n_Dynamic}
\end{align}

\subsection{Prediction}\label{Prediction}

Given measurements up to time $k \in [N_{n-1},N_n-1]$, trajectory predition for different $k+r$ is discussed as follows.

For $k+r \in [k+1,N_n-1]$, trajectory prediction is based on the following posterior density ($y_k=[x_k',x_{N_n}']'$) 
\begin{align}
&p(y_{k+r}|z^k)= \int p(y_{k+r}|y_k)p(y_k|z^k)dy_k\label{CM_Pred_1}
\end{align}
where the second term of the integrand is the output of the filter (see $\eqref{y_h}$--$\eqref{P_y_h}$), and the first term of the integrand is known by $\eqref{CM_n_Dynamic}$. So, for $k+r \in [k+1,N_n-1]$, the predicted state and its MSE matrix are obtained as
\begin{align}
&\hat{y}_{k+r|k}=G^y_{k+r|k}\hat{y}_{k}\label{y_k+n}\\
&\Sigma _{k+r|k}=B_{k+r|k}+G^y_{k+r|k}\Sigma _{k}(G^y_{k+r|k})'\label{S_k+n}
\end{align}
where $G^y_{k,k-1}=\left[
\begin{array}{cc}
G_{k,k-1} & G_{k,N_n}\\
0 & I
\end{array}\right]$ and 
\begin{align*}
G^y_{k+r|k}&=G^y_{k+r,k+r-1}G^y_{k+r-1,k+r-2} \cdots G^y_{k+1,k}\\
G^y_{k|k}&=I, \quad \forall k\\
B_{k+r|k}&=\sum _{i=k}^{k+r-1}G^y_{k+r|i+1}G^y_i(G^y_{k+r|i+1})'\\
\end{align*}
Then, the predicted estimate of $x_{k+r}$ and its MSE matrix are
\begin{align}
&\hat{x}_{k+r|k}=[I , 0]\hat{y}_{k+r|k}\label{x_k+n_2}\\
& P_{k+r|k}=[I , 0]\Sigma _{k+r|k}[I , 0]'\label{P_k+n_2}
\end{align}

For $k+r=N_n$, trajectory prediction is based on $p(x_{N_n}|z^k)$, which is available from the filter because it is a marginal of $p(x_k,x_{N_n}|z^k)$. We have
\begin{align}
&\hat{x}_{N_n|k}=[0 , I]\hat{y}_{k}\label{x_k+n_2_N}\\
& P_{N_n|k}=[0 , I]\Sigma _k[0 , I]'\label{P_k+n_2_N}
\end{align}

The $(n+1)$th waypoint is broadcast by $N_n$. So, the $(n+1)$th waypoint might be available at time $k$. Generally (even if later waypoints are known), for $k+r=N_q, n<q$, we have $p(x_{N_q}|z^{k})=\int p(x_{N_q}|x_{N_n})p(x_{N_n}|z^{k})dx_{N_n}$, where the first term of the integrand is given by $\eqref{D3}$ and the second term of the integrand is given by the filter (see $\eqref{N_filter_m}$--$\eqref{N_filter_C}$). Then,
\begin{align}
\hat{x}_{N_q|k}&=E[x_{N_q}|z^{k}]=\mu _{N_q}+G_{N_q,N_n}(\hat{x}_{N_n|k}-\mu _{N_n})\label{Nq|k}\\
P_{N_q|k}&=C_{N_q|k}=G_{N_q}+G_{N_q,N_n}P_{N_n|k}(G_{N_q,N_n})'\label{Nq|kC}
\end{align}
where $G_{N_q,N_n}=C_{N_q,N_n}(C_{N_n})^{-1}$ and $G_{N_q}=C_{N_q}-C_{N_q,N_n}(C_{N_n})^{-1}(C_{N_q,N_n})'$. 

If later waypoints (up to $(q+1)$th) are known, for $k+r \in [N_q+1,N_{q+1}-1], n\leq q$, we have ($y_k=[x_k' , x_{N_{q+1}}']'$) 
\begin{align}
&p(y_{k+r}|z^k)= \int p(y_{k+r}|y_{N_q})p(y_{N_q}|z^k)dy_{N_q}\label{CM_Pred_2}
\end{align}
where the first term of the integrand is known by ($y_k=[x_k',x_{N_{q+1}}']'$)
\begin{align}
y_k=
\left[
\begin{array}{cc}
G_{k,k-1} & G_{k,N_{q+1}}\\
0 & I
\end{array}\right] y_{k-1}+
e^y_{k-1}\label{CM_q+1_Dynamic}
\end{align}
The second term of the integrand of $\eqref{CM_Pred_2}$, $p(x_{N_q},$ $x_{N_{q+1}}|z^k)$, should be calculated. The terms $E[x_{N_q}|z^{k}]$, $C_{N_q|k}$, $E[x_{N_{q+1}}|z^k]$, and $C_{N_{q+1}|k}$ are obtained by $\eqref{Nq|k}$--$\eqref{Nq|kC}$. Also, similar to $\eqref{C_N2_N1|z}$, we have $C_{N_{q+1},N_q|k}=G_{N_{q+1},N_q}C_{N_q|k}=(C_{N_q,N_{q+1}|k})'$, where $G_{N_{q+1},N_q}=C_{N_{q+1},N_q}(C_{N_q})^{-1}$.

Given $p(x_{N_q},x_{N_{q+1}}|z^k)$ and $\eqref{CM_q+1_Dynamic}$, based on $\eqref{CM_Pred_2}$, we have
\begin{align}
&\hat{y}_{k+r|k}=G^y_{k+r|N_q}\hat{y}_{N_q|k}\label{y_k+n}\\
&\Sigma _{k+r|k}=B_{k+r|N_q}+G^y_{k+r|N_q}\Sigma _{N_q|k}(G^y_{k+r|N_q})'\label{S_k+n}
\end{align}
where $G^y_{k+r|N_q}=G^y_{k+r,k+r-1}G^y_{k+r-1,k+r-2} \cdots G^y_{N_q+1,N_q}$, $B_{k+r|N_q}=\sum _{i=N_q}^{k+r-1}G^y_{k+r|i+1}G^y_i$ $\cdot (G^y_{k+r|i+1})'$, and $G^y_{k|k}=I, \forall k$. Then, the predicted estimate at $k+r \in [N_q+1,N_{q+1}-1]$ is given by $\eqref{x_k+n_2}$--$\eqref{P_k+n_2}$, where $\hat{y}_{k+r|k}$ and $\Sigma _{k+r|k}$ are given by $\eqref{y_k+n}$--$\eqref{S_k+n}$.

\section{Simulations}\label{Simulation}

The CM sequence of Definition \ref{CM_Def} is simulated for modeling and prediction of trajectories with waypoints. Consider a two-dimensional scenario, where the state of an airliner at time $k$ is $x_k=[ \mathsf{x} , \mathsf{v}^{\mathsf{x}} , \mathsf{y} , \mathsf{v}^{\mathsf{y}}]_k'$, where $[\mathsf{x}_k,$ $\mathsf{y}_k]'$ is the position, and $[\mathsf{v}^{\mathsf{x}}_k,\mathsf{v}^{\mathsf{y}}_k]'$ is the velocity. Trajectories between four consecutive waypoints are simulated. Means and covariances of the states at waypoints and cross-covariances between the states at two consecutive waypoints are
\begin{align}
\mu _{N_1}&=[10000, 80, 5000, 30]'\label{m_1}\\
\mu _{N_2}&=[90000, 70, 30000, 50]'\label{m_2}\\
\mu _{N_3}&=[170000, 60, 170000, 60]'\label{m_3}\\
\mu _{N_4}&=[250000, 90, 200000, 30]'\label{m_4}\\
C_{N_i}&=\left[ \begin{array}{cccc}
10000 & 400 & 0 & 0\\\
400 & 100 & 0 & 0\\
0 & 0 & 10000 & 400\\
0 & 0 & 400 & 100
\end{array}\right], i=1,2,3,4\label{C}
\end{align}
\begin{align}
C_{N_i,N_{i-1}}&=\left[ \begin{array}{cccc}
8000 & 200 & 0 & 0\\\
200 & 70 & 0 & 0\\
0 & 0 & 8000 & 200\\
0 & 0 & 200 & 70
\end{array}\right], i=2,3,4\label{CC}
\end{align}

State evolution between waypoints ($\eqref{D1}$) is governed by a $CM_L$ model induced by a Markov model (Theorem \ref{CML_R_Dynamic_FQ_Proposition}). The corresponding Markov model is as follows. Consider Markov model $\eqref{Markov_Model}$ with
\begin{align}
M_{k+1,k}&=\text{diag}(F,F), \quad 
F=\left [\begin{array}{cc}
1 & T\\
0 & 1
\end{array}\right], \forall k\label{M1}\\
M_{k}&=\text{diag}(Q,Q), \quad Q=q\left[
\begin{array}{cc}
T^3/3 & T^2/2\\
T^2/2 & T
\end{array} \right]\label{M2}
\end{align} 
where $T=15$ seconds and $q=0.01$. The parameters of $\eqref{D1}$ are given by $\eqref{CML_Choice_1}$--$\eqref{CML_Choice_3}$. The waypoint times are $N_1=0$, $N_2=50$, $N_3=110$, and $N_3=150$. Also, the measurement equation is $z_k=Hx_k+v_k, k \in [1,N], H=\left[\begin{array}{cccc}
1 & 0 & 0 & 0\\
0 & 0 & 1 & 0
\end{array}\right]$, where $[v_k]_1^{N}$ ($\text{Cov}(v_k)=\text{diag}(100,100)$) is a zero-mean white NG sequence uncorrelated with $[x_k]$.

\begin{figure}
\centering
    \includegraphics[scale=0.45]{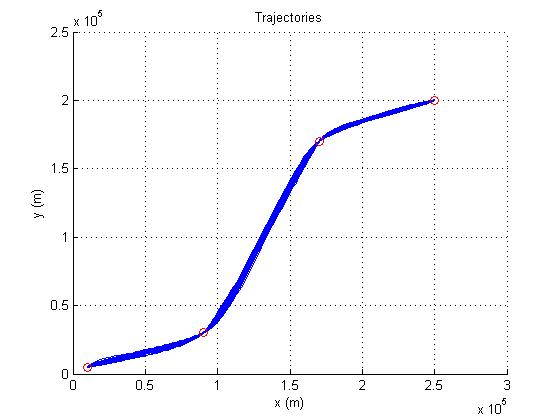}
\caption{Trajectories and waypoints.}
\label{F1}
\end{figure}

Fig. \ref{F1} shows several trajectories of the CM sequence governed by model $\eqref{D1}$ and $\eqref{D3}$ from the first to the fourth waypoint.

Assume measurements are available up to $k=4$ (the output of the filter is available at $k=4$). The goal is to predict the trajectory. Also, it is assumed that, in addition to the second waypoint, the third waypoint has already been broadcast and available. But the fourth waypoint is not known at $k=4$. As mentioned above, the evolution of the state between two consecutive waypoints is governed by a $CM_L$ model induced by the above Markov model. The joint endpoint distribution is an important part of a $CM_L$ model. Since there is no information about the fourth waypoint at $k=4$, it is natural to assume that the evolution of the state after the third waypoint is governed by the Markov model $\eqref{Markov_Model}$ with parameters $\eqref{M1}$--$\eqref{M2}$ and the initial distribution equal to the distribution at the third waypoint. This modeling assumption is well justified based on the definition of a $CM_L$ model induced by a Markov model (Section \ref{DD_Modeling}), as follows. Consider a Markov sequence governed by a Markov model $\eqref{Markov_Model}$. It is possible to obtain a $CM_L$ model governing this Markov sequence (i.e., the $CM_L$ model induced by the Markov model (Theorem \ref{CML_R_Dynamic_FQ_Proposition})). Assigning the right endpoint distribution to this induced $CM_L$ model, the corresponding $CM_L$ model (with its boundary conditions) governs the original Markov sequence, which is also governed by the original Markov model.     

To study the impact of a mismatch in the parameters, several mismatched cases are considered. The matched case, i.e., $\eqref{m_1}$--$\eqref{CC}$, is considered as case (i). The mismatched cases are:
\begin{itemize}
\item Case (ii):
\begin{align*}
\mu _{N_1}&=[10000, 60, 5000, 50]'\\
\mu _{N_2}&=[90000, 50, 30000, 70]'\\
\mu _{N_3}&=[170000, 40, 170000, 80]'\\
C_{N_i}&=\text{diag}(10^4,10^4,10^4,10^4), i=1,2,3\\
C_{N_i,N_{i-1}}&=\text{diag}(7000,6000,7000,6000), i=2,3
\end{align*}

\item Case (iii): Same as case (ii) except that $C_{N_i,N_{i-1}}=0, i =2,3$.

\end{itemize}

\begin{figure}
\centering
    \includegraphics[scale=0.5]{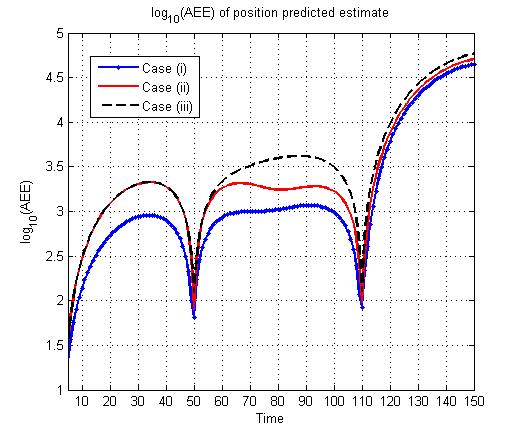}
\caption{Logarithm of AEE of position predictions ($\text{log}_{10}(\text{AEE}_{4+r|4})$).}
\label{F2}
\end{figure}

Fig. \ref{F2} shows the logarithm of the average Euclidean errors (AEE) \cite{Li_AEE} of the predictions of the position vector $[\mathsf{x}_k , \mathsf{y}_k ]'$ for cases (i)--(iii). Given measurements up to time $k$, the AEE of position prediction at time $k+r$ ($\text{AEE}_{k+r|k}$) is $\frac{1}{M}\sum _{i=1}^{M}\sqrt{(\mathsf{x}_{k+r}-\hat{\mathsf{x}}_{k+r|k})^2+(\mathsf{y}_{k+r}-\hat{\mathsf{y}}_{k+r|k})^2}$, where $[\mathsf{x}_{k+r} , \mathsf{y}_{k+r}]'$ is the true position at $k+r$ ($k+r = 5, \ldots , 150$) and $[\hat{\mathsf{x}}_{k+r|k} , \hat{\mathsf{y}}_{k+r|k}]'$ is its prediction using measurements up to time $k=4$, and $M=1000$ is the number of Monte Carlo runs. In cases (ii) and (iii), the means and covariances of the velocity at waypoints are highly mismatched. Case (iii) assumes that there is no correlation between states at different waypoints. In case (ii), the correlation coefficients between position components (and velocity components) at two consecutive waypoints are less than the true one. An underestimate of the correlation coefficient in case (ii) improves the prediction performance compared with case (iii) with zero correlation coefficient. Note that an overestimate of the correlation coefficient can degrade the performance, especially due to the mean and covariance mismatches at waypoints.    

Although it is reasonable to assume that the means of position components at waypoints are available, the impact of a mismatch in the means of position components in trajectory prediction is studied. The following mismatched cases are considered: 

\begin{itemize}

\item Case (iv): The differences with case (ii) are:
\begin{align*}
\mu _{N_1}&=[10500, 60, 5500, 50]'\\
\mu _{N_2}&=[90500, 50, 30500, 70]'\\
\mu _{N_3}&=[170500, 40, 170500, 80]'
\end{align*}

\item Case (v): Same as case (iv) except that $C_{N_i,N_{i-1}}=0, i = 2,3$.

\item Case (vi): The differences with case (iv) are:
\begin{align*}
C_{N_i}&=\text{diag}(10^5,10^4,10^5,10^4), i=1,2,3\\
C_{N_i,N_{i-1}}&=\text{diag}(70000,6000,70000,6000), i=2,3
\end{align*}

\item Case (vii): Same as case (vi) except that $C_{N_i,N_{i-1}}=0, i =2,3$.

\end{itemize}

\begin{figure}
\centering
    \includegraphics[scale=0.5]{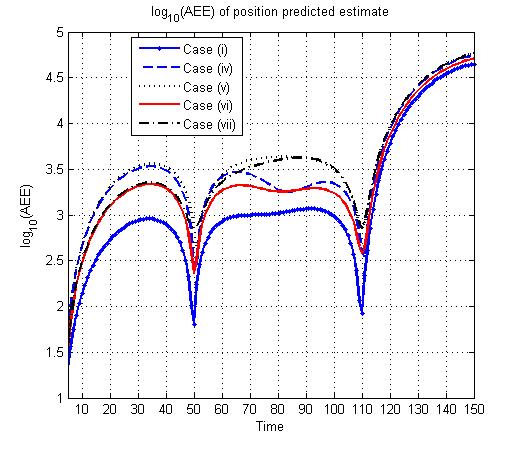}
\caption{Logarithm of AEE of position predictions ($\text{log}_{10}(\text{AEE}_{4+r|4})$).}
\label{F3}
\end{figure}

Fig. \ref{F3} shows the logarithm of the AEE of position predictions for case (i) and cases (iv) to (vii). The prediction performance degradation around the waypoints is due to the mismatch of the corresponding position means. A large covariance can compensate for the bias due to the mismatched mean. 

\section{Summary and Conclusions}\label{Summary}

Due to the air traffic control (ATC) regulations there are long range dependencies in trajectories of airliners. Such dependencies can be modeled by taking the waypoint information into account. In this paper, a conditionally Markov (CM) sequence has been proposed for modeling trajectories with waypoints. A dynamic model governing the proposed sequence has been presented. Filtering and trajectory prediction formulations have been obtained. First, the proposed CM sequence provides a simple and systematic approach for modeling trajectories with waypoints. Second, it is flexible to incorporate any kind of information available about the waypoints. Third, there is no restriction on the parameters of the presented dynamic model (this is good for analysis of the model and design of its parameters). Fourth, the presented dynamic model provides a systematic approach for reducing the uncertainty about the intent of an airliner as more measurements are received. It is based on calculation of the posterior state density at the next waypoint given measurement at the current time. Fifth, the presented dynamic model provides a systematic approach for handling inaccurate information about the waypoints (e.g., the state mean at a waypoint), based on appropriate covariance matrices. 

Suitable CM sequences can be systematically defined for trajectory modeling in different scenarios with waypoints and/or destination information available. The coorresponding dynamic models are simple and easy to apply. This is not necessarily the case about other stochastic sequences. For example, a generalization of the reciprocal sequence using the dynamic model of \cite{Levy_Dynamic} is not necessarily easy due to the structure of the model and its correlated dynamic noise \cite{White_Tracking2}.

More results about CM and reciprocal sequences can be found in \cite{}--\cite{}.

\subsubsection*{Acknowledgments}

Research was supported by NASA Phase03-06 through grant NNX13AD29A.

\end{document}